\documentclass[pra,twocolumn,showpacs,nobibnotes,floatfix,superscriptaddress]{revtex4}
\usepackage{latexsym,amsmath,amssymb,amsfonts,mathbbol,graphicx,color}
\usepackage{dcolumn}
\usepackage{bm}
\usepackage{graphicx}
\usepackage{epstopdf}

\begin{document}

\title{Encoding an Arbitrary State in a [7,1,3] Quantum Error Correction Code}

\author{Sidney D. Buchbinder}
\altaffiliation[Present address:]{ California Institute of Technology, Pasadena, CA 91125, USA}
\author{Channing L. Huang} 
\altaffiliation[Present address:]{ Princeton University, Princeton, NJ 08544, USA} 
\author{Yaakov S. Weinstein}
\affiliation{Quantum Information Science Group, {\sc Mitre},
260 Industrial Way West, Eatontown, NJ 07724, USA}

\begin{abstract}
We calculate the fidelity with which an arbitrary state can be encoded into a $[7,1,3]$ 
CSS quantum error correction code in a non-equiprobable Pauli operator error 
environment with the goal of determining whether this encoding can be used for 
practical implementations of quantum computation. This determination is accomplished by applying 
ideal error correction to the encoded state which demonstrates the correctability of 
errors that occurred during the encoding process. We then apply single-qubit Clifford 
gates to the encoded state and determine the accuracy with which these gates can be applied. 
Finally, fault tolerant noisy error correction is applied to the encoded states in the  
non-equiprobable Pauli operator error environment allowing us to compare noisy (realistic)
and perfect error correction implementations. We note that this maintains the 
fidelity of the encoded state for certain error-probability values. These results have 
implications for when non-fault tolerant procedures may be used in practical quantum 
computation and whether quantum error correction should be applied at every step in a 
quantum protocol. 
\end{abstract}

\maketitle

\section{introduction}
Quantum fault tolerance is a framework designed to allow accurate implementations of quantum algorithms despite inevitable
errors \cite{G,ShorQFT,Preskill,AGP}. While the construction of this framework is important and necessary to demonstrate the
possibility of successful quantum computation, the realization of a quantum computer which utilizes the complete edifice of 
quantum fault tolerance requires huge numbers of qubits and is a monumental task. Thus, it is worthwhile to explore whether it 
is possible to achieve successful quantum computation without using the full toolbox of quantum fault tolerance techniques. 
Initial studies along these lines have been done for logical zero encoding in the Steane [7,1,3] 
quantum error correction code \cite{YSW}. In this paper we extend these results by determining the accuracy with which 
an arbitrary state can be encoded into the Steane code in a non-equiprobable Pauli operator error environment. 
Encoding an arbitrary state in the Steane code cannot be done in a fault tolerant manner \cite{Preskill}, and 
we explore whether the accuracy achieved for a non-fault tolerant encoding is sufficient for use in a realistic 
quantum computation. We then apply (noisy) single qubit Clifford gates and error correction to this encoded 
state, again concentrating on the accuracy of the implementation.

The first step in any fault tolerant implementation of quantum computation is to encode the necessary quantum information
into logical states of a quantum error correction (QEC) code \cite{book,ShorQEC}. Here we make use of the [7,1,3] Steane 
code \cite{CSS,Steane} which can completely protect one qubit of quantum information by encoding the information 
into seven physical qubits. Encoding information into the 
Steane code can be done via the gate sequence originally designed in \cite{Steane}. However, this method is not 
fault tolerant as an error on a given qubit may spread to other qubits. A fault tolerant method exists for encoding 
only the logical zero and one states \cite{Preskill}. Here we study the accuracy of the arbitrary state encoding 
in attempt to determine whether it could be useful for practical implementations of quantum computation. We assume a 
non-equiprobable error model \cite{QCC,AP} in which qubits taking part in any gate are subject to a
$\sigma_{x}$ error with probability $p_{x}$, a $\sigma_{y}$ error with probability $p_{y}$, and a $\sigma_{z}$ error 
with probability $p_{z}$. Thus, an attempted implementation of a single qubit transformation $T$ on qubit $j$ described by 
density matrix $\rho_0$ would produce:
\begin{equation}
\rho=\displaystyle\sum\limits_{a}^{0,x,y,z} p_{a}\sigma_{a}^{j}T^{j}\rho_{0}{T^{j}}^{\dag}{\sigma_{a}^{j}}^{\dag},
\label{TransformEquation}
\end{equation}
where $\sigma_{0}$ is the identity matrix, and $p_{0}=1-\displaystyle\sum\limits_{i=x,y,z} p_{i}$. 
A two-qubit controlled-\textsc{not} (\textsc{cnot}) gate with control qubit $j$ and target qubit $k$ 
($\textsc{c}_{j}\textsc{not}_{k}$) would cause the following evolution of an initial two-qubit 
density matrix $\rho_0$:
\begin{equation}
\rho=\displaystyle\sum\limits_{a,b}^{0,x,y,z} p_{a}p_{b}\sigma_{a}^{j}\sigma_{b}^{k}\textsc{c}_{j}\textsc{not}_{k}\rho_{0}\textsc{c}_{j}\textsc{not}_{k}^{\dag}{\sigma_{b}^{k}}^{\dag}{\sigma_{a}^{j}}^{\dag}.
\label{CNOT}
\end{equation}

After noisy encoding of an arbitrary state, we apply ideal error correction as a means of determining whether the 
errors that occurred during encoding can, at least in principal, be corrected, thus making the encoded state useful 
for practical instantiations of quantum computation. In addition, we apply each of three logical Clifford gates 
(Hadamard, \textsc{not}, phase) to the noisy encoded state, and measure the accuracy of their implementation 
in the non-equiprobable error environment. We apply ideal 
error correction after each of these gates. Finally, we apply noisy QEC to the encoded states, to model 
the reliability with which data can be encoded and maintained in a realistic noisy environment.

In this work, we utilize three different measures of fidelity to evaluate accuracy. 
The first fidelity quantifies the accuracy of the seven-qubit state after the intended operations have 
been performed compared to the desired seven-qubit state. 
The second fidelity measures the accuracy of the logical qubit, the single qubit of quantum information stored in 
the QEC code. To obtain this measure, we perfectly decode the seven-qubit state and partial trace over the six 
(non-logical) qubits.  
The final fidelity measure compares the state of the seven qubits after application of perfect error correction 
to the state that has undergone perfect implementation of the desired transformations. 
This fidelity reveals whether the errors that occur in a noisy encoding process can be corrected for use in 
fault tolerant quantum computation.

\section{General State Encoding}

\begin{figure}[t]
\includegraphics[width=8cm]{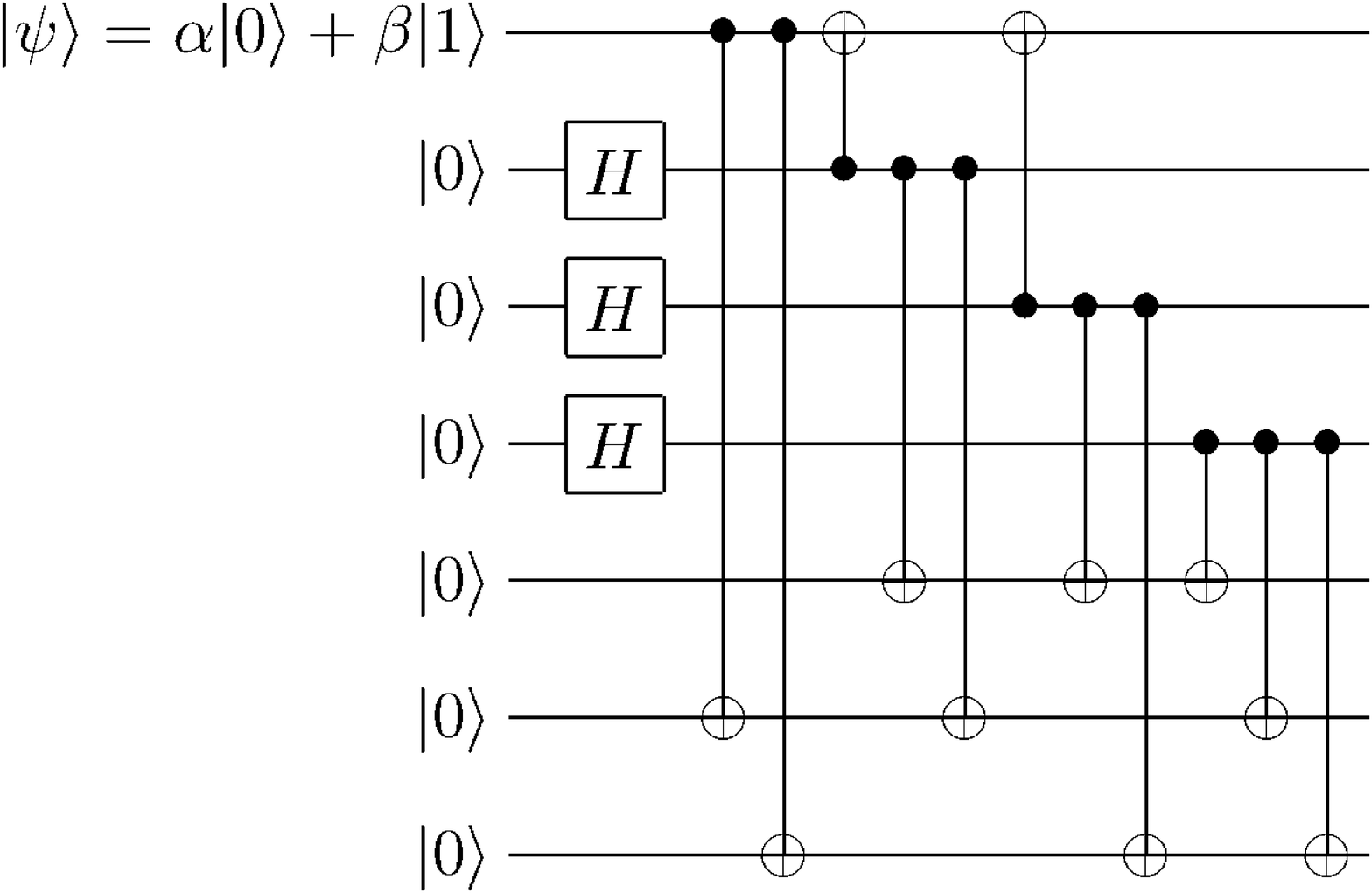}
\caption{Gate circuit for  encoding an arbitrary state into the $[7,1,3]$ Steane QEC code. 
$H$ represents a single qubit Hadamard gate, and vertical lines connect the control ($\bullet$) and target
($\oplus$) qubits of a \textsc{cnot} gate.
}
\label{Encoding}
\end{figure}

Our first step is to encode an arbitrary pure state 
$|\psi(\alpha,\beta)\rangle = \cos\alpha|0\rangle+e^{i\beta}\sin\alpha|1\rangle$ 
into a [7,1,3] QEC code using the gate sequence shown in Fig.~\ref{Encoding}. 
As each gate is applied, the non-equiprobable error environment causes the qubits taking 
part in the gate to probabilistically undergo Pauli errors, as in Eqs.~\ref{TransformEquation} 
and \ref{CNOT}. The resulting state of the noisy, error-prone [7,1,3] encoding, $\rho^{(7)}_{Enc_N}$, depends 
on the state being encoded and the probabilities of error ($p_{x}$, $p_{y}$, and $p_{z}$). 
We determine the accuracy of the encoding process by calculating the fidelity as compared to an 
ideally (non error-prone) encoded state $\rho^{(7)}_{Enc_I}$. The seven-qubit fidelity, 
$Tr[\rho^{(7)}_{Enc_N}\rho^{(7)}_{Enc_I}]$, reflects the accuracy with which the general state encoding process is
achieved and is given (to second order) in the appendix, Eq.~\ref{FidEnc7}. 
This expression reveals that no first order error terms are dependent on $\beta$, and the only first order 
term dependent on the state to be encoded is $p_{z}$. This indicates a relatively small dependence on initial 
state in general. Encoding with an initial state of $\alpha=0$ or $\alpha=\frac{\pi}{2}$ results in the highest 
fidelity, while encoding an intial state with $\alpha=\frac{\pi}{4}$ results in a lower fidelity. We further note that 
there is no dominant error term, in that the magnitudes of the coefficients of the first order terms are similar. 

\begin{figure}[t]
\centering
\includegraphics[width=4cm]{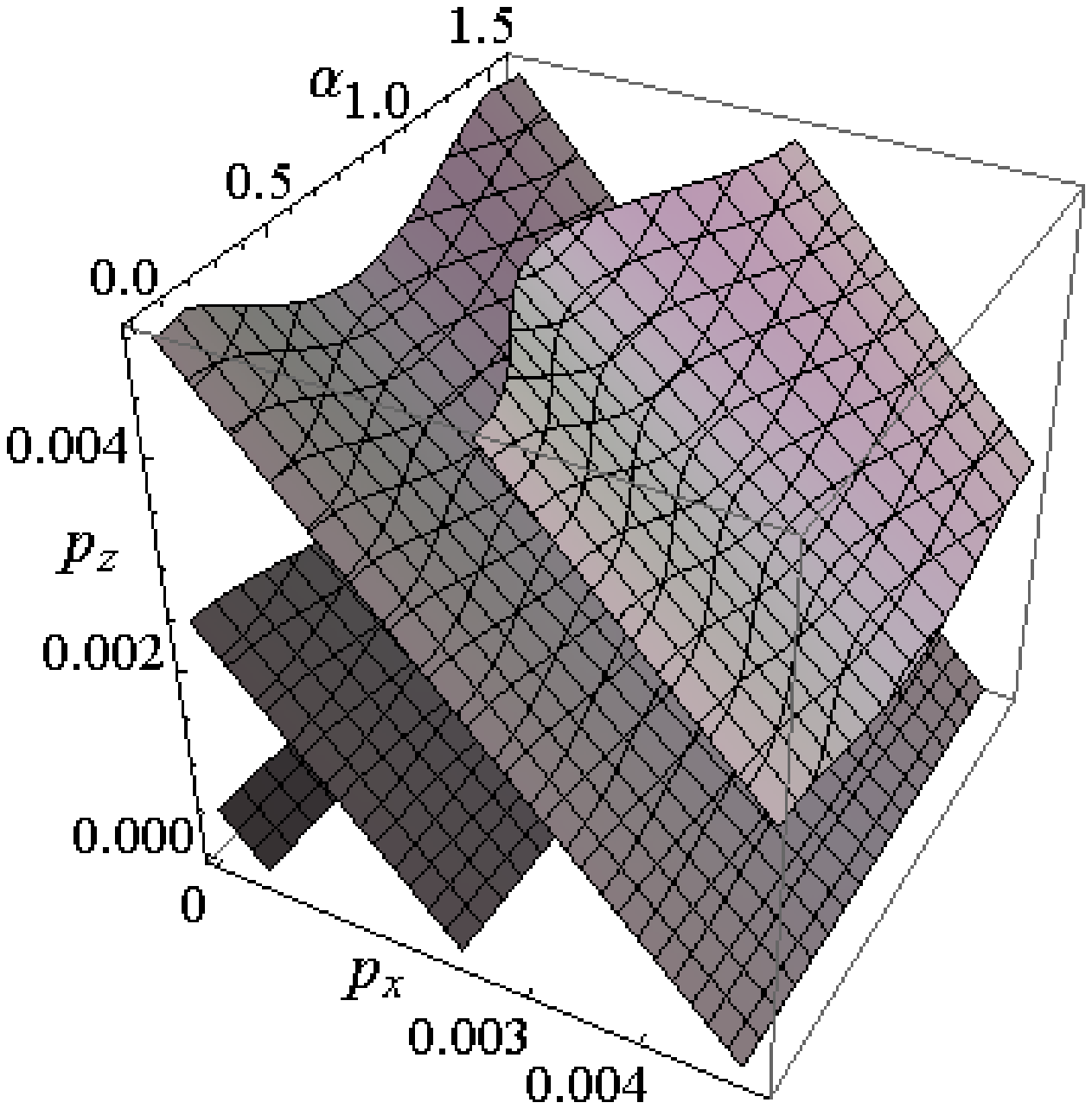}
\includegraphics[width=4cm]{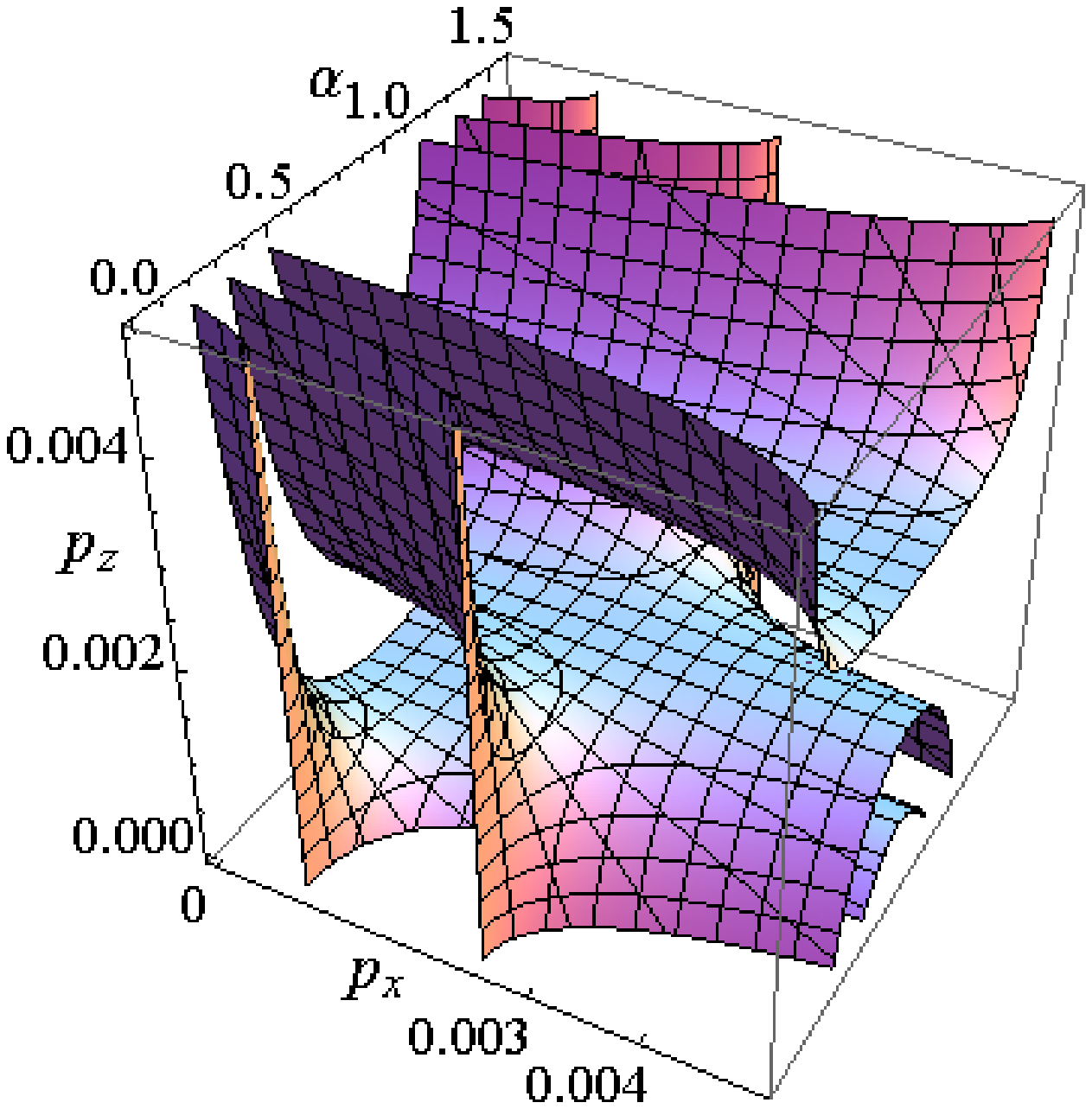}
\caption{Contour plots for the fidelity of the arbitrary state $[7,1,3]$ encoding procedure as a 
function of initial state parameterized by $\alpha$ ($\beta$ is set to 0), and error probabilities
$p_{x}$, and $p_{z}$ ($p_{y}$ is set to zero). 
Left: The seven qubit fidelity for the general encoding procedure. 
Right: The single data qubit fidelity for the general encoding procedure. 
The contours for the seven qubit fidelity values are (from top to bottom) 
0.85, 0.90, 0.95, and 0.99. The contours for the single qubit fidelities 
are (from top to bottom) 0.97, 0.985, and 0.995.}
\label{GeneralFidelity}
\end{figure}

The seven-qubit fidelity quantifies the accuracy with which the seven physical qubits are in the 
desired state. However, certain errors may not impact the single 
logical qubit of quantum information stored in the 
QEC code. We would like to determine the accuracy of that single 
logical qubit of information. To do this we perfectly decode the seven-qubit 
system, and partial trace over qubits 2 through 7. We compare the resulting one-qubit state 
$\rho_{Enc_N}^{(1)}$ to the starting state of the logical qubit. This fidelity is then given by:
$\langle\psi(\alpha,\beta)|\rho_{Enc_N}^{(1)}|\psi(\alpha,\beta)\rangle$,
and is given (to second order) in Eq.~\ref{FidEnc1}. 
We find that this fidelity is highly dependent on the initial state. All terms, including the first order terms, 
are dependent on $\alpha$. In addition, dependence on $\beta$ appears in all terms but $p_{z}$ and $p_{z}^{2}$. 
As above, no single type of error dominates the loss of fidelity, as indicated by the similar coefficients of the first 
order probability terms. The initial states that result in the highest fidelity occur at $\alpha=0$ and 
$\alpha=\frac{\pi}{2}$, while the state that results in the lowest fidelity occurs at 
$\alpha=\frac{\pi}{4},\beta=\frac{\pi}{4}$. Contour plots of the seven qubit and single 
qubit fidelities are shown in Fig.~\ref{GeneralFidelity}.

\section{Application of Clifford Gates}
We now apply the one logical qubit Clifford gates to the encoded arbitrary state using the non-equiprobable error model 
described above, and determine the accuracy of their implementation. CSS codes in general, and the 
Steane code in particular, allow logical Clifford gates to be implemented bitwise, so that only one (physical) qubit 
gates are required. The Clifford gates applied in this study are the Hadamard (H), \textsc{not} (X), 
and phase (P) gates:
\begin{eqnarray*}
H = \frac{1}{\sqrt{2}}\left(
\begin{array}{cc}
1 & 1 \\
1 & -1 \\
\end{array}
\right),\;\;
X = \left(
\begin{array}{cc}
0 & 1 \\
1 & 0 \\
\end{array}
\right),\;\;
P = \left(
\begin{array}{cc}
1 & 0 \\
0 & i \\
\end{array}
\right).
\end{eqnarray*}
The logical Hadamard gate is implemented by applying a single qubit Hadamard gate on each of the seven 
qubits in the encoding. The logical \textsc{not} gate is implemented by applying a single qubit 
\textsc{not} gate to the first three qubits of the encoded state. The logical phase gate is implemented 
by applying an inverse phase gate to each of the seven qubits. Each of these gates is applied to the 
noisily encoded arbitrary state of the last section in the non-equiprobable error environment 
such that a qubit acted upon by a gate evolves via Eq.~\ref{TransformEquation}. 
We quantify the accuracy of the gate implementation by calculating the fidelities as a comparison of 
the state after application of the Clifford gate with the state of a noisily encoded arbitrary state that 
undergoes perfect application of the Clifford transformation: $Tr[\rho^{(7)}_{Enc_N C_I}\rho^{(7)}_{Enc_N C_N}]$ and 
$Tr[\rho^{(1)}_{Enc_N C_I}\rho^{(1)}_{Enc_N C_N}]$, where $C$ represents $H$, $X$, or $P$. Plots of these 
fidelities are given in Fig.~\ref{Cliffordfidelity}, and the expressions of these fidelities 
are given in the appendix, Eqs.~\ref{Had7qubitfidelity} - \ref{Phase1qubitfidelity}.

\begin{figure}[t]
\centering
\includegraphics[width=4cm]{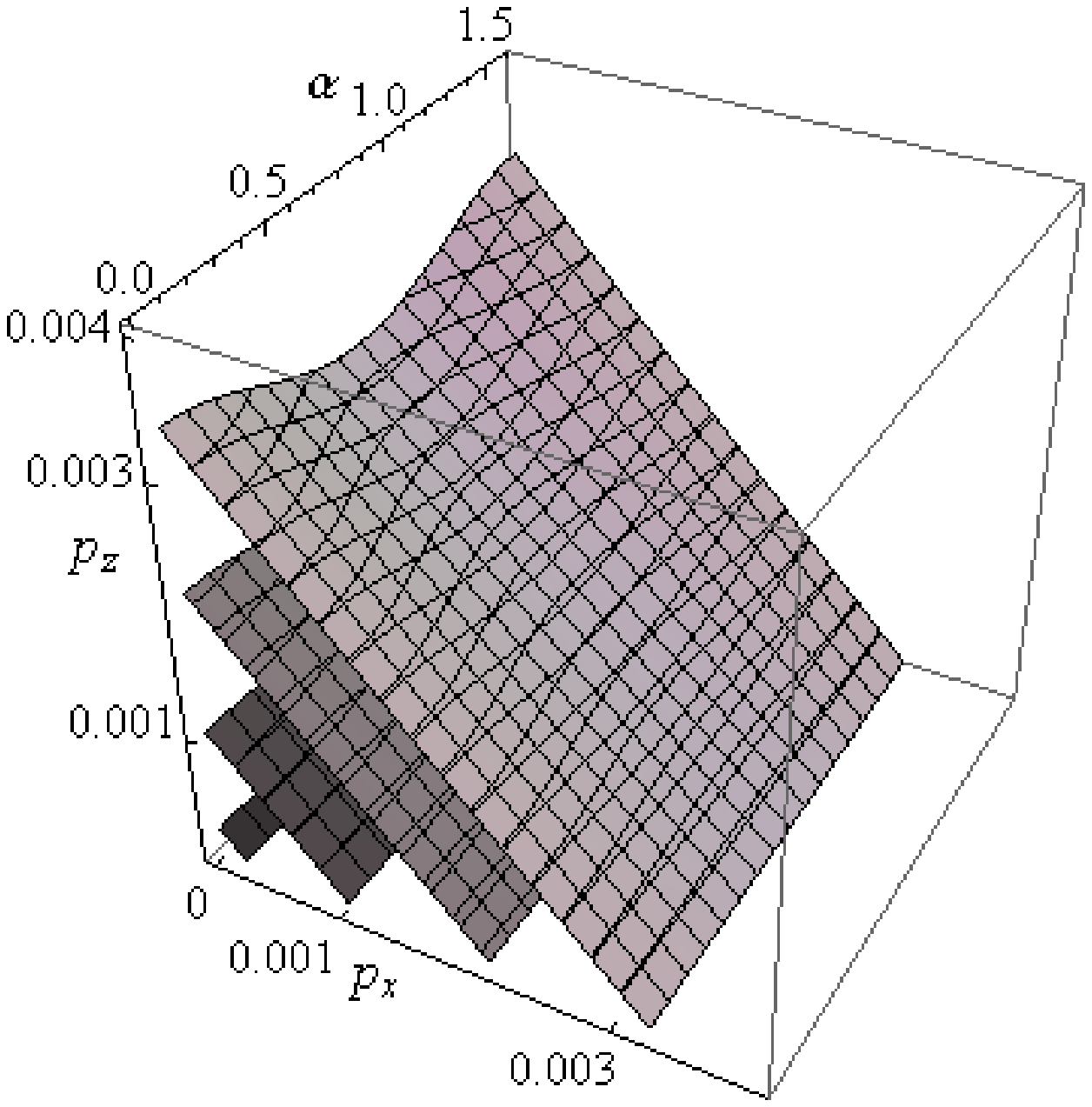}
\includegraphics[width=4cm]{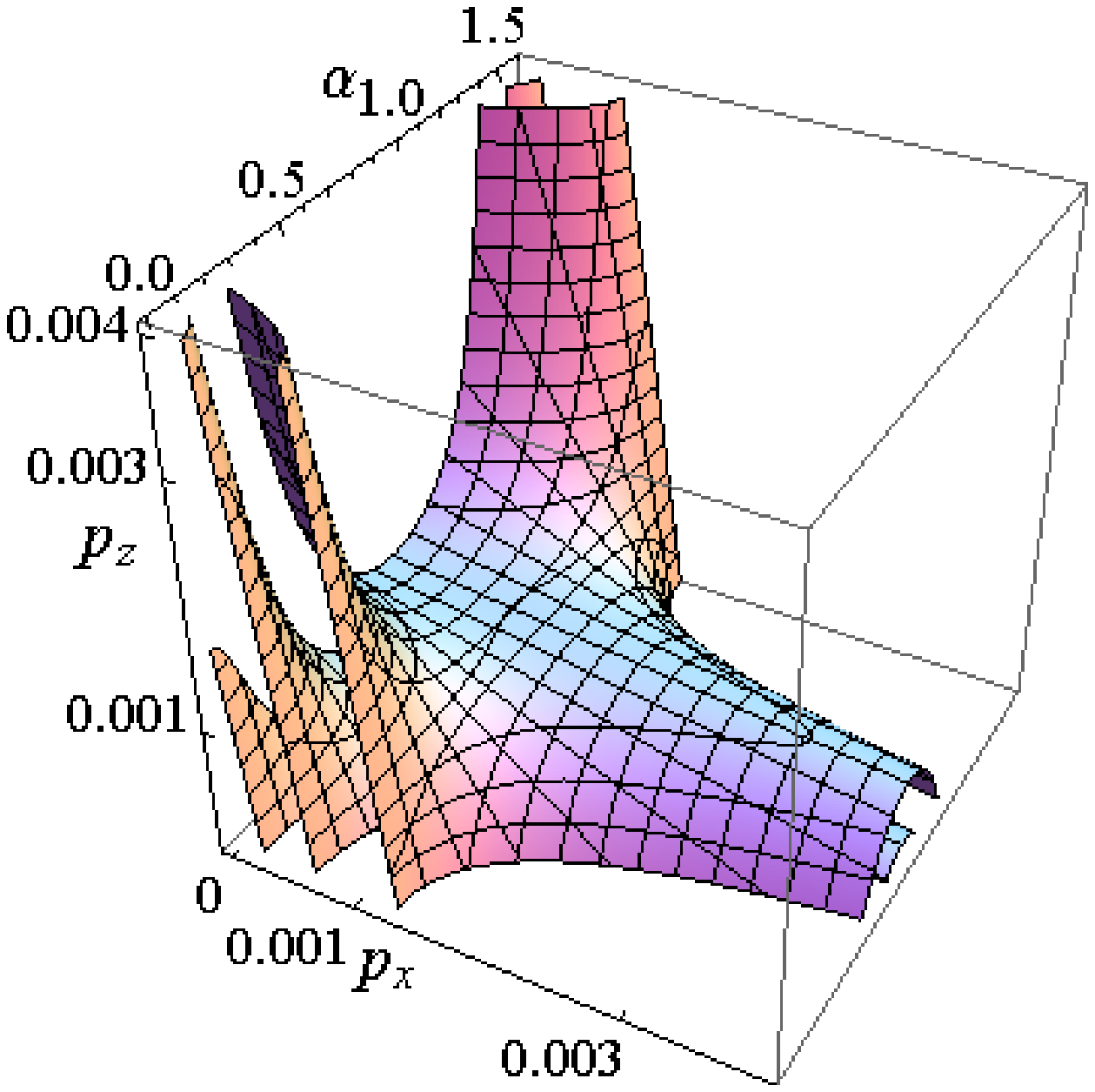}
\includegraphics[width=4cm]{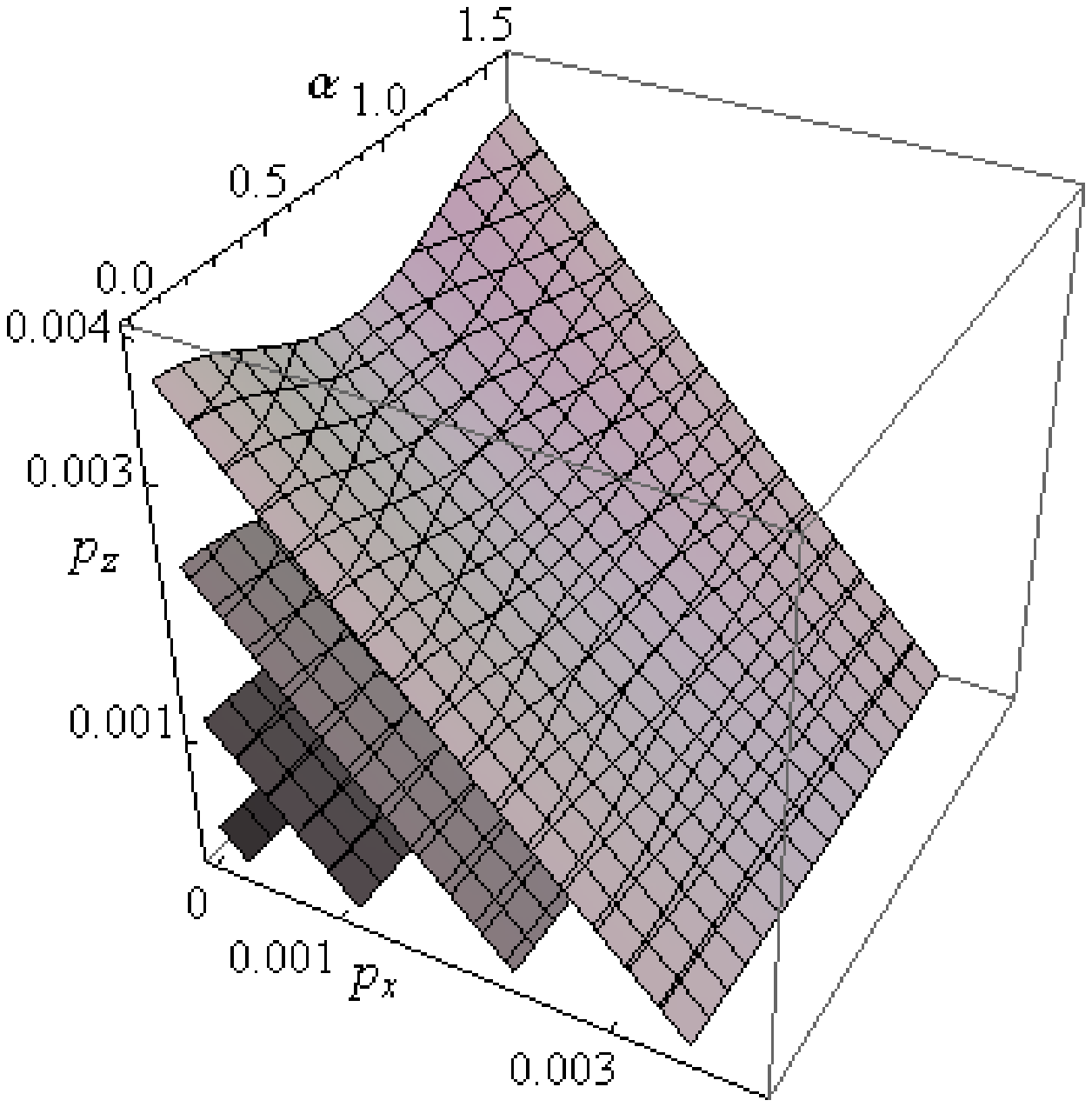}
\includegraphics[width=4cm]{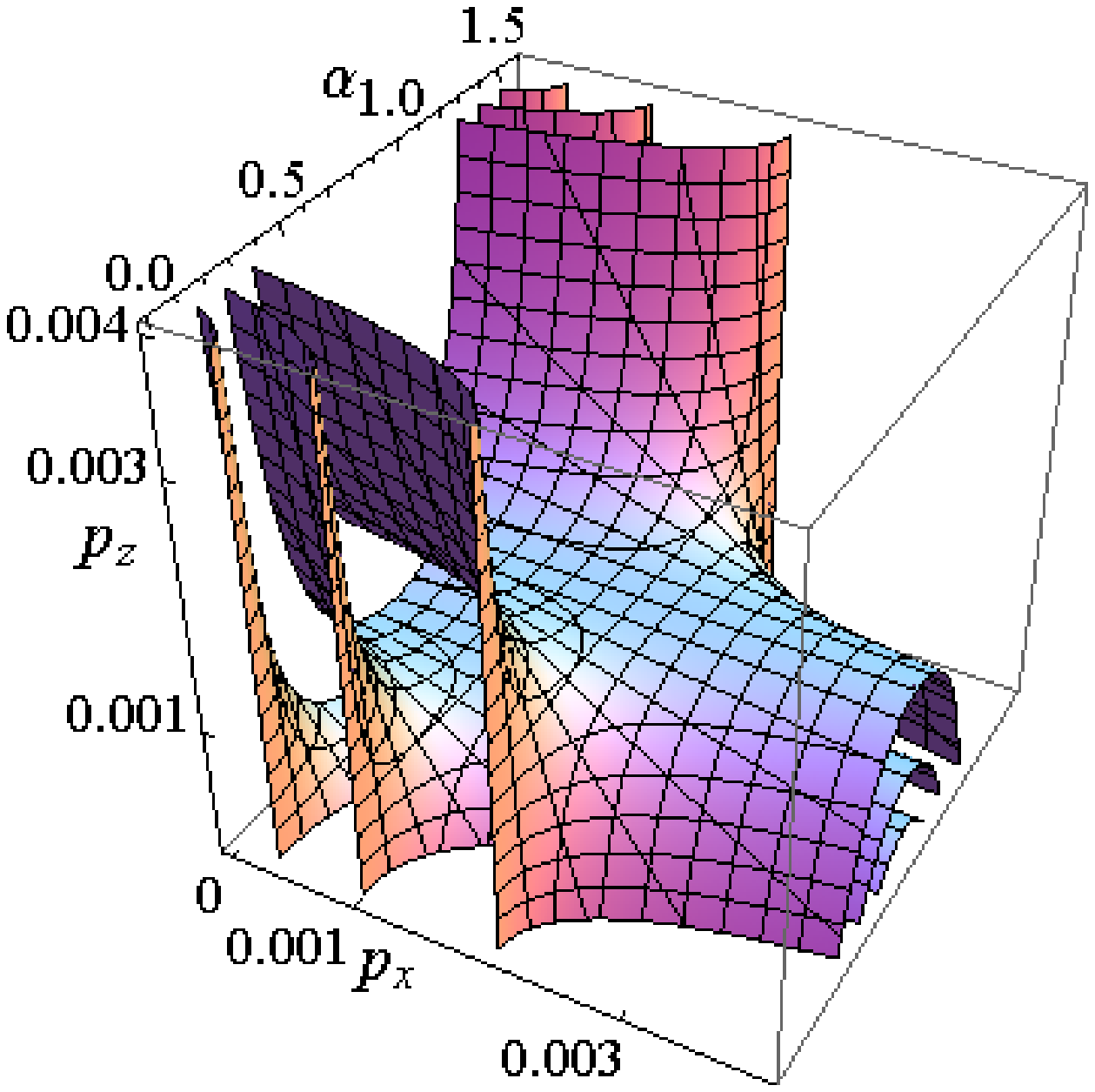}
\includegraphics[width=4cm]{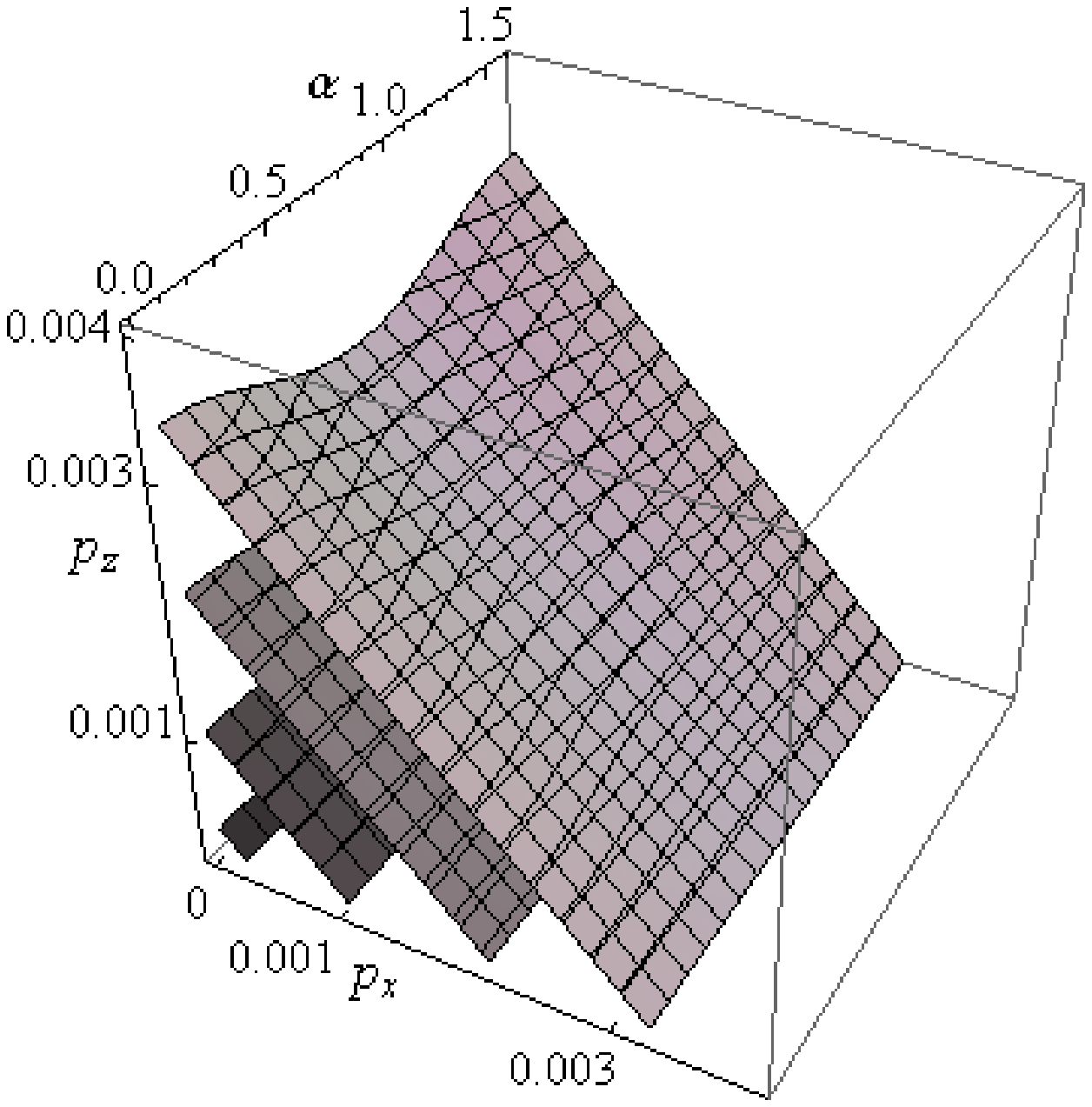}
\includegraphics[width=4cm]{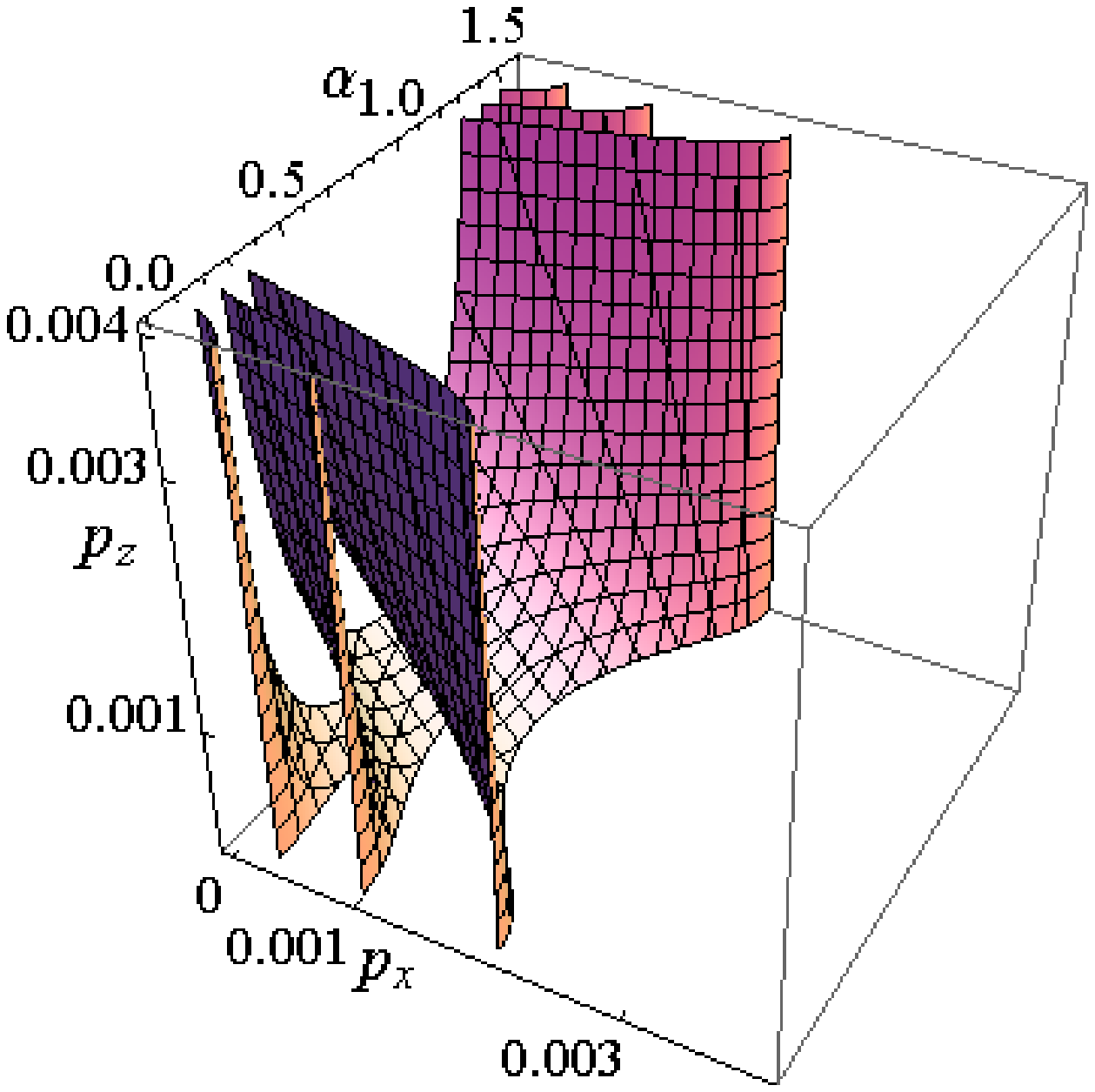}
\caption{Contour plots of fidelity for application of the single qubit Clifford gates on an 
arbitrary state encoded in the $[7,1,3]$ code as a function of the initial state paramaterized by 
$\alpha$ ($\beta$ is set to 0), and error probabilities $p_{x}$, and $p_{z}$ ($p_{y}$ is set to zero). 
Left: Seven qubit fidelities, 
Right: single logical qubit fidelities for the Hadamard, \textsc{not}, and phase gates (top to bottom). 
The contours for the seven qubit fidelity values are (from top to bottom) 0.85, 0.90, 0.95, and 0.99. 
The contours for the single qubit fidelities are (from top to bottom) 0.97, 0.985, and 0.995. 
Note that the contour plots of the seven qubit fidelities for the Hadamard and phase gates 
(top-left and bottom-left) are nearly indistinguishable.}
\label{Cliffordfidelity}
\end{figure}

The encoded state (seven qubit) fidelities for the three Clifford gates follow several general trends. 
The fidelities of the Hadamard and phase gates, Eqs.~\ref{Had7qubitfidelity} and 
\ref{Phase7qubitfidelity}, are nearly identical and slightly lower than the fidelity of the 
\textsc{not} gate, Eq.~\ref{Not7qubitfidelity}. 
This is likely due to the fact that the same number of gates are applied to implement the 
logical Hadamard and phase gates which is more than needed to implement the logical \textsc{not} 
gate. In addition, all three fidelities have similar dependence on the initial state as 
paramaterized by $\alpha$ and $\beta$ (though dependence on $\beta$ is small). 
While the coefficient of the first order $p_{y}$ term is larger than the coefficients 
of the first order $p_{x}$ and $p_{z}$ terms for all three fidelities, 
the difference is only slight. No specific error dominates the loss in fidelity.

Different trends are apparent in the fidelities of the single logical qubit. First, the fidelity of 
the Hadamard state, Eq.~\ref{Had1qubitfidelity}, is lower than the fidelities of the \textsc{not} 
and phase states, Eq.~\ref{Not1qubitfidelity} and \ref{Phase1qubitfidelity}. Furthermore, all three 
fidelities exhibit relatively large dependence on initial state, in that $\alpha$ appears in every 
first order term, and $\beta$ appears in nearly all first order terms. The fidelities after application 
of the \textsc{not} gate and Hadamard gate exhibit similar magnitude dependences on $\alpha$, while 
the phase fidelity changes more significantly as the initial state varies. For all gates, the most 
stable initial states are $\alpha=0, \alpha=\frac{\pi}{2}$, 
in that a higher fidelity occurs at the same $p_{x}$, $p_{y}$, and $p_{z}$ values when compared with 
fidelities of other initial states. We note as well that the single data qubit fidelities of the \textsc{not} 
gate and phase gate become independent of $\sigma_z$ errors for these values of $\alpha$.
In both fidelity measures there are first order terms which would 
ideally be suppressed to second order through the application of quantum error correction.

\section{Quantum Error Correction}
The purpose of encoding a state via the $[7,1,3]$ code is to protect the data from being lost due to 
errors. While errors may have occurred during the encoding itself, these can be corrected by subsequent 
application of QEC if the encoding is done in a fault tolerant fashion thus ensuring that at most one 
data qubit will be corrupted. In this paper we have encoded an arbitrary state in necessarily non-fault
tolerant fashion. Nevertheless, we would like to know if QEC applied after encoding can suppress errors in 
the same way as would be done for a fault tolerant encoding.

\subsection{Perfect Error Correction}
We apply perfect (non error-prone) error correction to the noisily encoded arbitrary state 
and to the encoded states after one of the three Clifford gates have been applied. Performing 
ideal error correction on noisy states enables us to determine if there is any feasible recovery 
from the errors that occur during the noisy encoding process and noisy gate application. Perfectly 
applied error correction should suppress the probability of error terms in the fidelity to at least 
second order for the gate sequence encoding method to be usable in practical quantum computing. 
If the first order error probability terms are not suppressed under ideal error 
correction conditions, error correction that occurs in a noisy environment (which is more realistic) 
will certainly not be able to reliably preserve encoded data.

\begin{figure}[t]
\includegraphics[width=8cm]{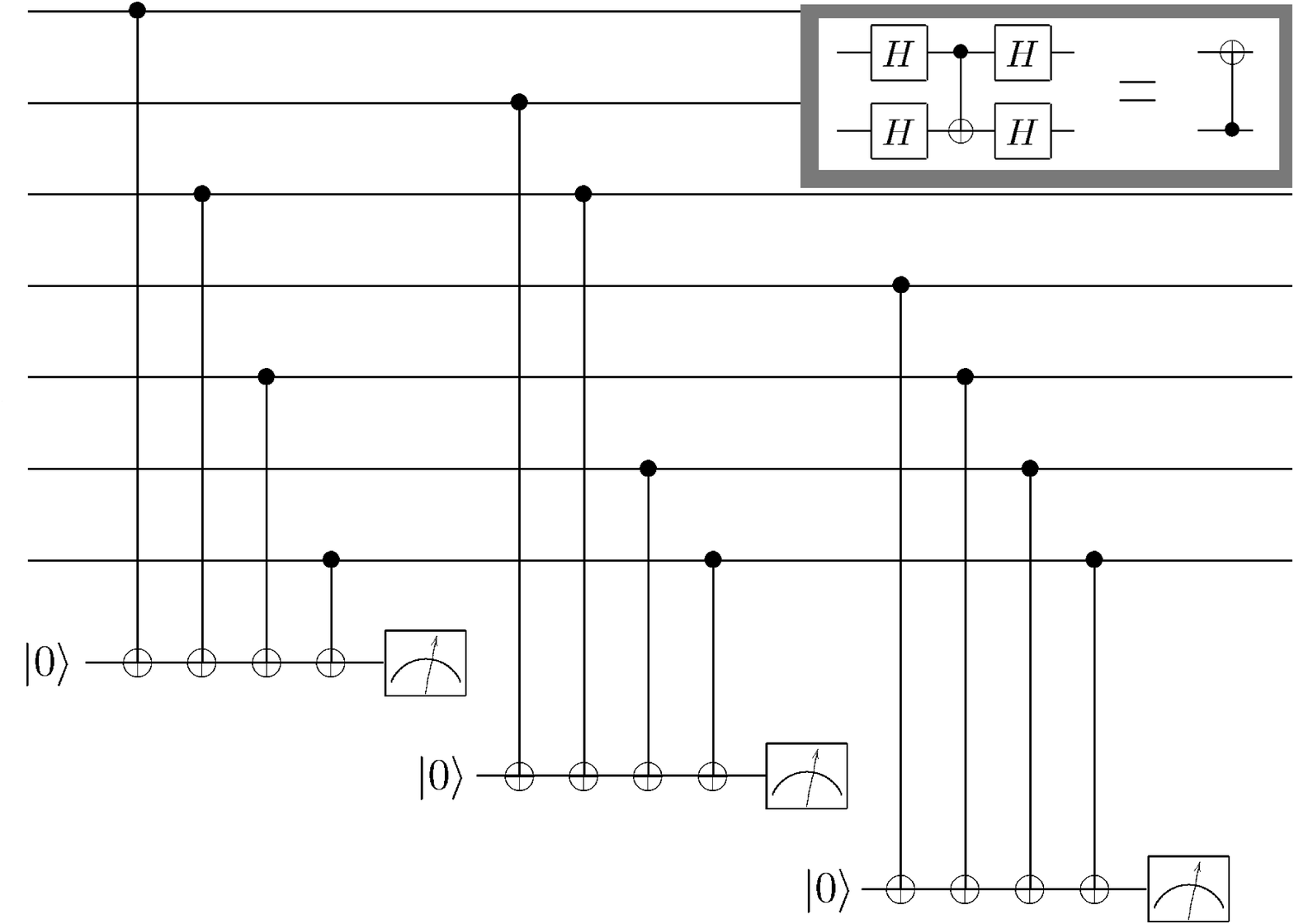}
\caption{Main: Non-fault tolerant syndrome measurement for the $[7,1,3]$ code. Applying these three 
syndrome measurements will reveal the presence of bit flips. Subsequently applying Hadamard gates to 
the data qubits and applying the same three syndrome measurements will reveal the presence of phase 
flips. 
Box: A useful equality that allows us to simplify the error correction procedure. We can avoid 
implementing Hadamard gates by reversing the control and target qubits of the \textsc{cnot} gates.}
\label{SimplifiedQEC}
\end{figure}

The fidelity resulting from implementing perfect error correction on the encoded arbitrary state is given by 
Eq. \ref{PQEC}. While perfect error correction on the encoded arbitrary state does improve the fidelity, 
there is a remaining first order term $p_{z}$ (the errors associated with the first order 
$p_{x}$ and $p_{y}$ have been suppressed by perfect error correction). This would suggest that the gate
sequence encoding scheme is not appropriate for practical quantum computation. However, for certain 
initial states $\alpha = 0,\frac{\pi}{2}$ the first order error term drops out and thus this process 
can be used to create logical $|0\rangle$ and $|1\rangle$ states.

We observe similar trends when perfect QEC is performed on the states that have undergone Clifford transformations, 
Eqs. \ref{PQECHad}, \ref{PQECNot}, and \ref{PQECPhase}. In all three cases, the error probabilities in 
the fidelities are suppressed to second order in $p_{x}$ and $p_{y}$, while $p_{z}$ is suppressed to second 
order only for the initial states $|0\rangle$ and $|1\rangle$. This implies that this encoding 
process and the application of a logical one-qubit Clifford gates can be used for practical quantum computation
only for the initial states $|0\rangle$ and $|1\rangle$.

\subsection{Noisy Error Correction}
\begin{figure}[b]
\includegraphics[width=6cm]{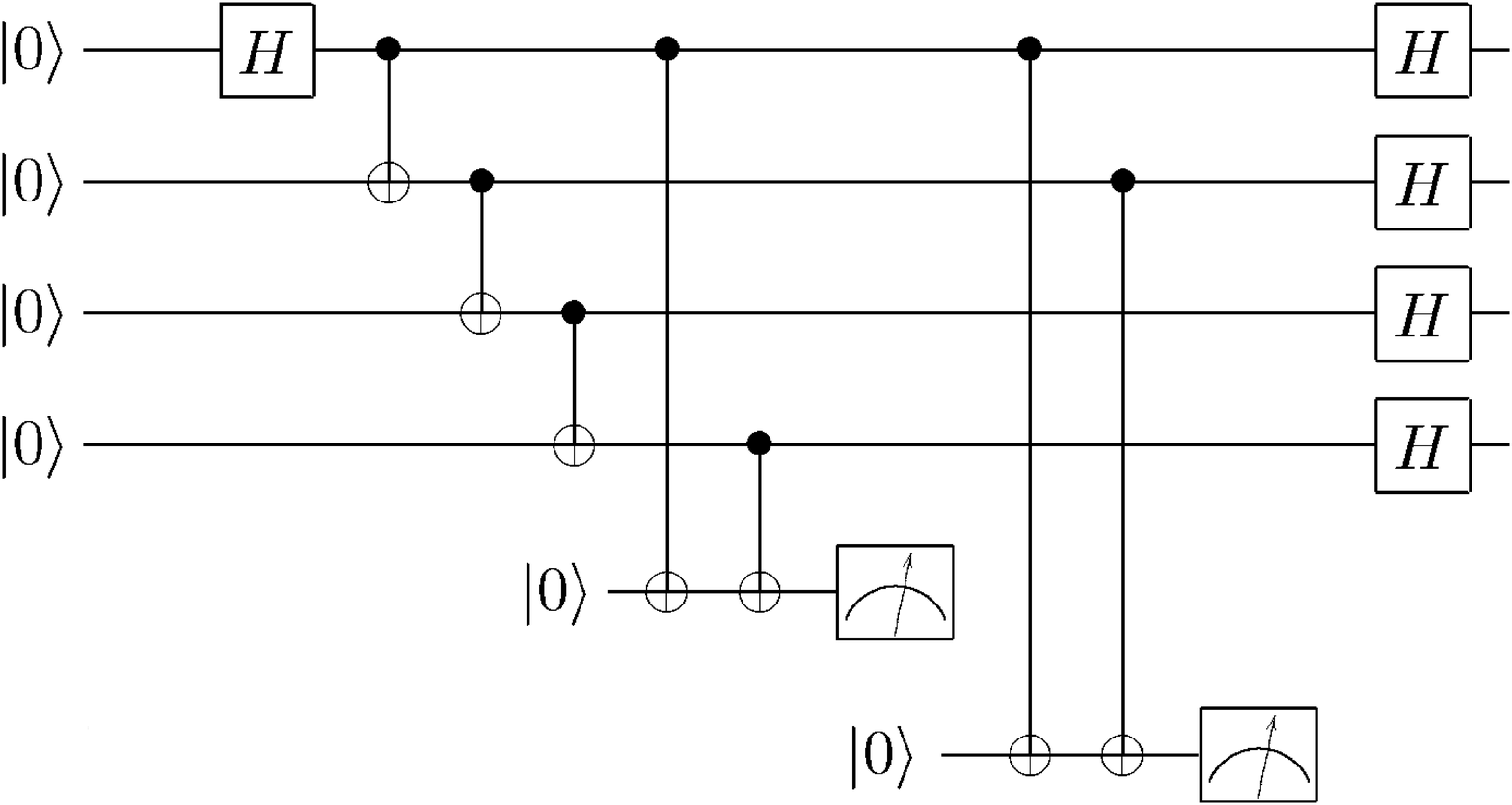}
\includegraphics[width=9cm]{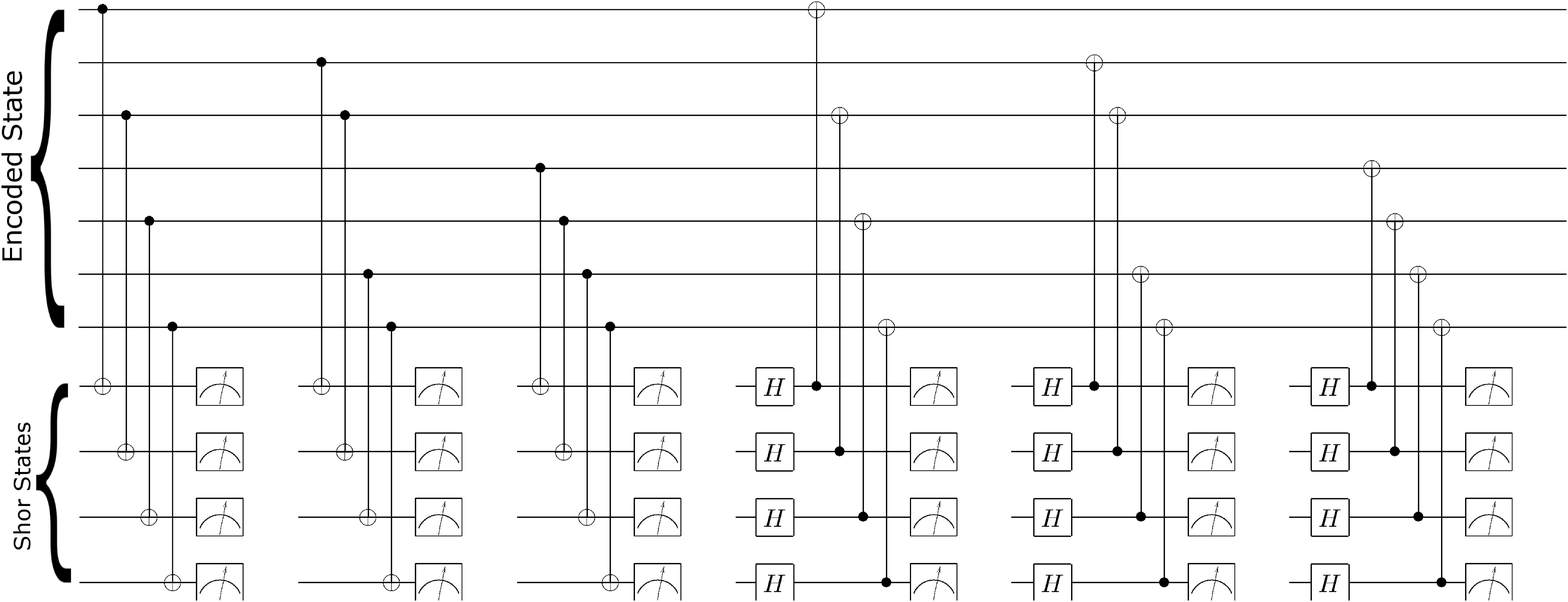}
\caption{Top: Preparation of the Shor state. To construct the Shor state, we first prepare a GHZ state. 
To ensure fault tolerant construction, we perform two parity checks between qubits 1 and 4 and qubits 1 and 2. 
If the state is not correct, it is discarded and a new Shor state is prepared. Application of Hadamard 
rotations to the four qubits completes the preparation of the Shor states.
Bottom: The fault tolerant Steane code utilizing Shor states to perform the syndrome measurements. 
To check for bit flips, three sets of \textsc{cnot}s are performed with the 
Shor state qubits as the target. The parity of the ancilla qubits 
determines the syndrome value. To perform phase flip checks, we utilize the equality shown in the 
Box of Fig.~\ref{SimplifiedQEC}. We apply Hadamard rotations the the ancilla bits, cancelling the Hadamards
at the end of the Shor state construction, and flip the direction of the \textsc{cnot}s. Measuring in the 
$x$-basis allows us to eliminate the final Hadamards on the ancilla qubits. }
\label{FullQEC}
\end{figure}

Real error correction will be noisy. Thus, we apply a fault tolerant error correction scheme in the 
non-equiprobable error environment to the arbitrary encoded state to determine what might occur in a more 
realistic quantum computation. In adhering to the rules of fault tolerance we utilize Shor states as 
syndrome qubits (Fig.~\ref{FullQEC}). The Shor states themselves are prepared in a noisy environment 
such that the Hadamard evolution is properly described by Eq.~\ref{TransformEquation}, and the evolution
of the \textsc{cnot} gates is described by Eq.~\ref{CNOT}. We verify the Shor state by performing two 
parity checks (also done in the non-equiprobable error environment). If the state is not 
suitable, it is thrown away and a new state is prepared. After applying the necessary \textsc{cnot}s between 
the data and ancilla qubits, we measure the ancilla qubits the parity of which determines the syndrome value. 
The phase flip syndrome measurements are performed in a similar manner (Fig.~\ref{FullQEC}). 
We analyze only the case where all four ancilla qubits are measured as zero for each of the syndrome measurements. 
We apply each syndrome check twice to confirm the correct parity measurement, as errors may occur while 
implementing the syndrome measurements themselves.

\begin{figure}[t]
\includegraphics[width=4cm]{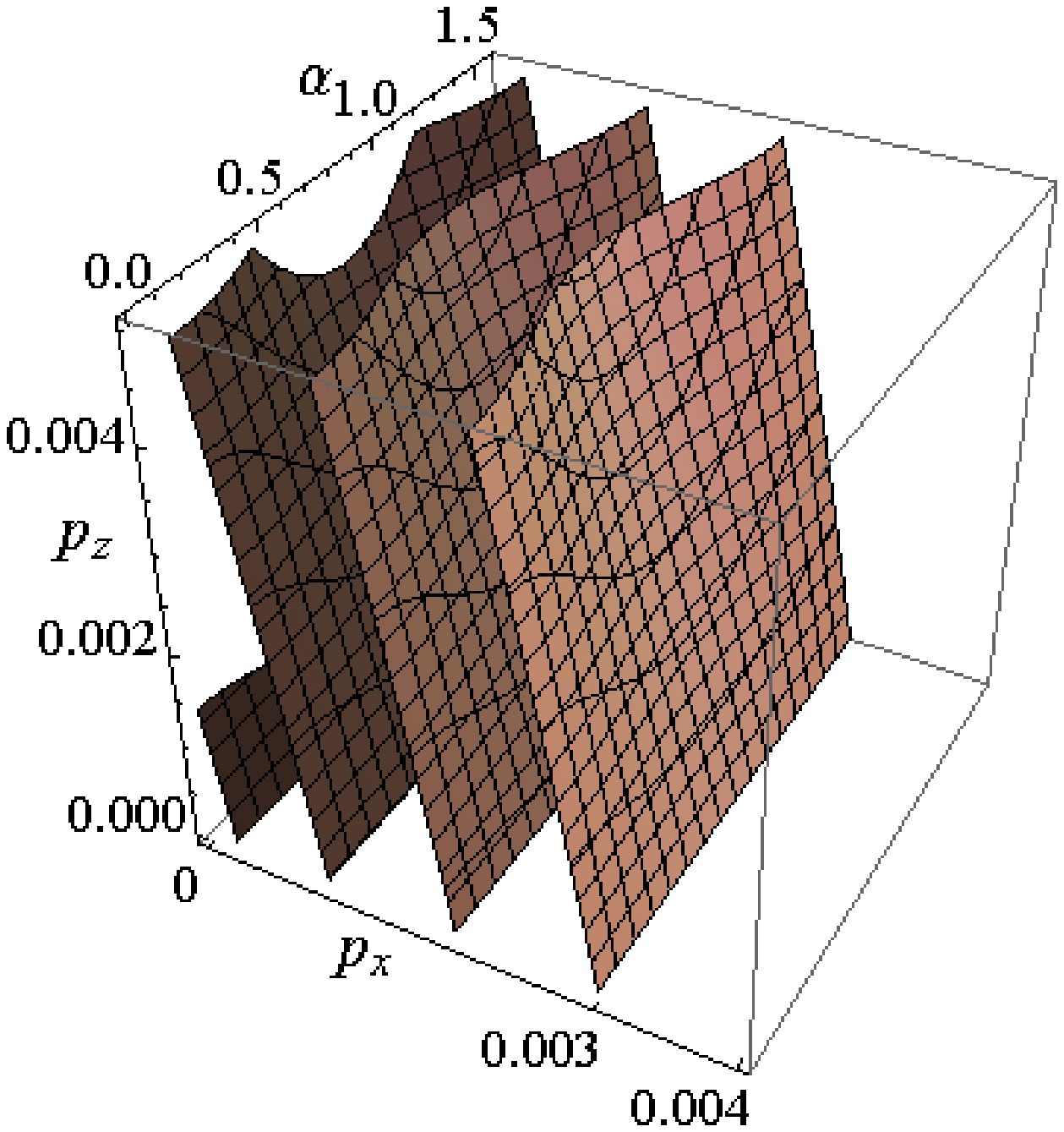}
\includegraphics[width=4cm]{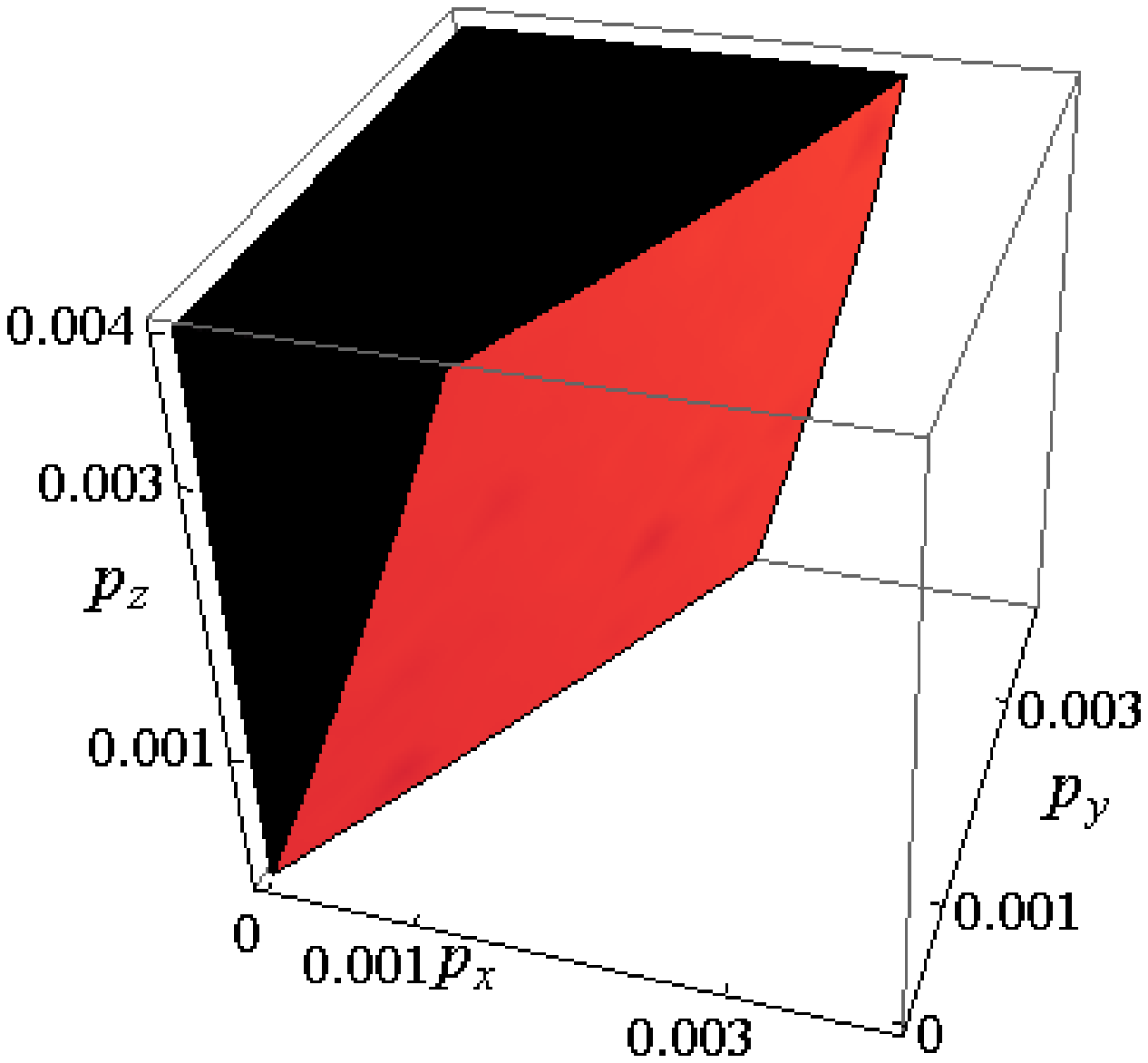}
\caption{Left: Contour plots of seven qubit fidelity for noisy QEC on an arbitrary state encoded in the $[7,1,3]$ 
code as a function of the initial state paramaterized by $\alpha$ ($\beta$ is set to 0), $p_{x}$, and 
$p_{z}$ ($p_{y}$ is set to zero. The contours for the fidelity values are (from top to bottom) 0.85, 
0.90, 0.95, and 0.99. 
Right: The shaded region indicates where in probability space the seven qubit fidelity of a noisily encoded arbitrary 
initial state increases after applying QEC. We find that the $\alpha$ and $\beta$ of the initial state
do not significantly affect this region. In this plot $\alpha$ and $\beta$ are both set to $0$.}
\label{FIGQEC}
\end{figure}

If error correction is to perform as desired and maintain an encoded state with high reliability, 
the fidelity measurement of the post error-prone QEC state should be higher than the fidelity of 
the noisily encoded state with no error correction applied. The fidelity should furthermore 
be greater than the fidelity expected for unencoded states, which would indicate that going through 
the [7,1,3] encoding process and QEC procedure effectively maintains information more reliably than 
undergoing no encoding at all. The fidelity for the (noisily) error corrected arbitrary encoded 
state is given in Eq.~\ref{QEC}.

We find that the fidelity after an error prone error correction code is quite comparable to the 
fidelity of the encoded state prior to error correction. This fidelity still includes first order 
terms in $p_{x}$, $p_{y}$, and $p_{z}$. There is little dependence on the initial state, as $\alpha$ 
only appears in the $p_{z}$ term, though the initial states with $\alpha=0$ and $\alpha=\frac{\pi}{2}$ 
result in slightly higher fidelities than other initial states. We note that the $p_{x}$ term is now 
dominant. Furthermore, while the $p_{x}$ term in the fidelity of the error-prone QEC state is higher than 
the corresponding term in the fidelity of the pre-QEC encoded state, Eq.~\ref{FidEnc7}, the $p_{y}$ and 
$p_{z}$ terms of the QEC state are both significantly lower than the corresponding terms of the encoded 
state fidelity. Thus, in cases where $p_{x}$ is lower than $p_{y}$ and $p_{z}$, the noisy error 
correction scheme does indeed improve fidelity. However, when the $p_{x}$ is high, the noisy error 
correction can significantly lower the fidelity of the encoded state. Fig.~\ref{FIGQEC} displays a 
contour plot of the seven qubit fidelity with noisy error correction, and a plot showing the region 
in probability-space for which the post-QEC states attain a higher fidelity than the noisily encoded 
states (prior to error correction).

The fidelity of the single logical qubit for the QEC code is given in Eq.~\ref{QEC1}. We find that the single logical qubit fidelity after error-prone QEC is comparable to the fidelity of the single logical qubit before QEC was applied. This fidelity contains first order terms in $p_x$, $p_y$, and $p_z$. Furthermore, the fidelity is highly dependent on initial state, as $\alpha$ appears in all first order terms. The initial states $\alpha=0$ and $\alpha=\frac{\pi}{2}$ 
result in higher fidelities than other initial states. We note that in this measure of fidelity, the $p_x$ error is dominant as well.

When noisy error correction is implemented a second time on the system, the fidelities remain nearly the 
same, as seen in Eqs.~\ref{QEC2X} and~\ref{QEC2X1}. 

The seven-qubit fidelity of the state after applying noisy error correction to a noisy arbitrary encoded 
state that has undergone a logical \textsc{not} gate is given by Eq.~\ref{QECNot}. This fidelity 
features similar characteristics to the fidelity when noisy error correction is applied to the general 
encoded state, Eq.~\ref{QEC}. It contains first order terms, with the $p_{x}$ term again being dominant. 
The single logical qubit fidelity of this state is given by Eq.~\ref{QECNot1}, and bears strong resemblance 
to the post QEC single qubit fidelity given by Eq.~\ref{QEC1}.
These results indicate that it may not necessary or beneficial to 
perform error correction after every step in a quantum procedure, but rather that one should apply QEC only at 
specific intervals or after specific gate sequences. 

\section{Conclusion}
In conclusion, we have explicitly evaluated the accuracy with which an arbitrary state can be encoded 
into the Steane $[7,1,3]$ QEC code. The calculated fidelities for the direct gate encoding method are only 
slightly worse than those calculated for the fault tolerant encoding of the zero state \cite{YSW}. We then 
applied logical Clifford gates to the encoded arbitrary state. The fidelity of the seven qubit system is 
primarily dependent on the number of gates applied, while the fidelity of the single data qubit is dependent 
on both the specific gate being applied and on the initial state being encoded. Ideal error correction is 
able to suppress some of the error probability terms in the fidelity, though first order $p_{z}$ error 
terms are still present for most initial states. Applying noisy quantum error correction maintains the 
fidelity of the state at a level comparable to that prior to QEC application, though the relative values 
of $p_{x}$, $p_{y}$, and $p_{z}$ can cause significant increases or decreases in this fidelity. Our results
indicate that the [7,1,3] QEC gate sequence encoding is only practical for specific initial states. They 
also indicate that QEC need not be applied after every step in a quantum protocol. We believe
that simulations of this sort will help determine what quantum gates will be acceptable in order to 
successfully implement quantum computation.

\begin{widetext}
\appendix
\section{Fidelities}

In this appendix we present the fidelities discussed in the main part of the paper. All fidelities are given 
to second order. In the equations, $\alpha$ and $\beta$ paramaterize the initial state of the qubit to be encoded, 
$|\psi(\alpha,\beta)\rangle = \cos\alpha|0\rangle+e^{i\beta}\sin\alpha|1\rangle$, $p_{x}$ represents the 
probability of a $\sigma_{x}$ error, $p_{y}$ represents the probability of a $\sigma_{y}$ error, and $p_{z}$ 
represents the probability of a $\sigma_{z}$ error. The superscript label of each equation indicates 
the number of qubits in the state for which the fidelity is calculated: the entire seven physical qubit 
system or the single logical qubit.

Eq. \ref{FidEnc7} is the fidelity of the seven qubit encoding of an arbitrary state: 
\begin{eqnarray}
F^{(7)}_{enc}&=&1-22 p_x-25 p_y-(23-2 \text{Cos}[4 \alpha ]) p_z+\left(\frac{515}{2}-\frac{3}{2} \text{Cos}[4 \alpha ]+3 \text{Cos}[2 \beta ] \text{Sin}[2 \alpha ]^2\right) p_x^2\nonumber\\
&+&\left(532+2 \text{Cos}[4 \alpha ]-2 \text{Cos}[2 \beta ] \text{Sin}[2 \alpha ]^2\right) p_x p_y+\left(\frac{1205}{4}-\frac{1}{4} \text{Cos}[4 \alpha ]+\frac{1}{2} \text{Cos}[2 \beta ] \text{Sin}[2 \alpha ]^2\right) p_y^2\nonumber\\
&+&(486-42 \text{Cos}[4 \alpha ]) p_x p_z+\left(\frac{1127}{2}-\frac{93}{2} \text{Cos}[4 \alpha ]\right) p_y p_z+\left(\frac{577}{2}-\frac{81}{2} \text{Cos}[4 \alpha ]\right) p_z^2.
\label{FidEnc7}
\end{eqnarray}

Eq. \ref{FidEnc1} is the fidelity of the single logical qubit of an encoded arbitrary state that has undergone perfect (non error-prone) decoding:
\begin{eqnarray}
F^{(1)}_{enc}&=&1-\left(\frac{9}{2}+\frac{3}{2} \text{Cos}[4 \alpha ]-3 \text{Cos}[2 \beta ]\text{Sin}[2 \alpha ]^2\right) p_x-\left(\frac{15}{2}-\frac{3}{2} \text{Cos}[4 \alpha ]+\text{Cos}[2 \beta ]\text{Sin}[2 \alpha ]^2\right) p_y-(5-5 \text{Cos}[4 \alpha ]) p_z\nonumber\\
&+&\left(\frac{45}{2}+\frac{15}{2} \text{Cos}[4 \alpha ]-15\text{Cos}[2 \beta ]\text{Sin}[2 \alpha ]^2\right) p_x^2+(53+7 \text{Cos}[4 \alpha ]-46\text{Cos}[2 \beta ]\text{Sin}[2 \alpha ]^2) p_x p_y\nonumber\\
&+&\left(\frac{103}{2}-\frac{43}{2} \text{Cos}[4 \alpha ]+17\text{Cos}[2 \beta ]\text{Sin}[2 \alpha ]^2\right) p_y^2+(28-28 \text{Cos}[4 \alpha ]-56\text{Cos}[2 \beta ]\text{Sin}[2 \alpha ]^2) p_x p_z\nonumber\\
&+&(82-82 \text{Cos}[4 \alpha ]+16\text{Cos}[2 \beta ]\text{Sin}[2 \alpha ]^2) p_y p_z+(45-45 \text{Cos}[4 \alpha ]) p_z^2.
\label{FidEnc1}
\end{eqnarray}

Eqs. \ref{Had7qubitfidelity} - \ref{Phase1qubitfidelity} are the fidelities of the state after application of a noisy Clifford gate 
compared to the noisily encoded arbitrary state with ideal applciation of the Clifford gate. Eqs.~\ref{Had7qubitfidelity}, 
\ref{Not7qubitfidelity}, and \ref{Phase7qubitfidelity} represent the fidelities of the encoded seven-qubit system 
after a logical Hadamard, \textsc{not}, and phase gate has been applied, respectively. 
Eqs. \ref{Had1qubitfidelity}, \ref{Not1qubitfidelity}, and \ref{Phase1qubitfidelity} represent the fidelities 
of the single logical qubit after a logical Hadamard, \textsc{not}, and phase gate has been applied, respectively:

\begin{eqnarray}
F^{(7)}_{H}&=&1-51 p_x-57 p_y-(53-4 \text{Cos}[4 \alpha ]) p_z+\left(1403-6 \text{Cos}[4 \alpha ]+12 \text{Cos}[2 \beta ] \text{Sin}[2 \alpha ]^2\right) p_x^2\nonumber\\
&+&\left(2882+12 \text{Cos}[4 \alpha ]-8 \text{Cos}[2 \beta ] \text{Sin}[2 \alpha ]^2\right) p_x p_y+\left(1640-\text{Cos}[4 \alpha ]+2 \text{Cos}[2 \beta ] \text{Sin}[2 \alpha ]^2\right) p_y^2\nonumber\\
&+&\left(2732-198 \text{Cos}[4 \alpha ]+4 \text{Cos}[2 \beta ] \text{Sin}[2 \alpha ]^2\right) p_x p_z+\left(\frac{6047}{2}-\frac{431}{2} \text{Cos}[4 \alpha ]-\text{Cos}[2 \beta ] \text{Sin}[2 \alpha ]^2\right) p_y p_z\nonumber\\
&+&(1543-194 \text{Cos}[4 \alpha ]) p_z^2
\label{Had7qubitfidelity}
\end{eqnarray}

\begin{eqnarray}
F^{(1)}_{H}&=&1-\left(\frac{39}{2}+\frac{9}{2} \text{Cos}[4 \alpha ]-12\text{Cos}[2 \beta ]\text{Sin}[2 \alpha ]^2\right) p_x-\left(\frac{117}{4}-\frac{9}{4} \text{Cos}[4 \alpha ]-\frac{1}{2}\text{Cos}[2 \beta ]\text{Sin}[2 \alpha ]^2\right) p_y\nonumber\\
&-&\left(\frac{65}{4}-\frac{53}{4} \text{Cos}[4 \alpha ]-\frac{3}{2} \text{Cos}[2 \beta ]\text{Sin}[2 \alpha ]^2\right) p_z+(453+99 \text{Cos}[4 \alpha ]-348\text{Cos}[2 \beta ]\text{Sin}[2 \alpha ]^2) p_x^2\nonumber\\
&+&(1094+154 \text{Cos}[4 \alpha ]-760\text{Cos}[2 \beta ]\text{Sin}[2 \alpha ]^2) p_x p_y+\left(\frac{1663}{2}-\frac{259}{2} \text{Cos}[4 \alpha ]-31\text{Cos}[2 \beta ]\text{Sin}[2 \alpha ]^2\right) p_y^2\nonumber\\
&+&(526-382 \text{Cos}[4 \alpha ]-740) p_x p_z+(985-829 \text{Cos}[4 \alpha ]-134\text{Cos}[2 \beta ]\text{Sin}[2 \alpha ]^2) p_y p_z\nonumber\\
&+&\left(\frac{849}{2}-\frac{837}{2} \text{Cos}[4 \alpha ]-87\text{Cos}[2 \beta ]\text{Sin}[2 \alpha ]^2\right) p_z^2
\label{Had1qubitfidelity}
\end{eqnarray}

\begin{eqnarray}
F^{(7)}_{X}&=&1-47 p_x-53 p_y-(49-4 \text{Cos}[4 \alpha ]) p_z+\left(\frac{2441}{2}-\frac{15}{2} \text{Cos}[4 \alpha ]+15 \text{Cos}[2 \beta ] \text{Sin}[2 \alpha ]^2\right) p_x^2\nonumber\\
&+&\left(2465+13 \text{Cos}[4 \alpha ]-8 \text{Cos}[2 \beta ] \text{Sin}[2 \alpha ]^2\right) p_x p_y+\left(1414-\text{Cos}[4 \alpha ]+2 \text{Cos}[2 \beta ] \text{Sin}[2 \alpha ]^2\right) p_y^2\nonumber\\
&+&(2260-184 \text{Cos}[4 \alpha ]) p_x p_z+(2604-202 \text{Cos}[4 \alpha ]) p_y p_z+(1365-174 \text{Cos}[4 \alpha ]) p_z^2
\label{Not7qubitfidelity}
\end{eqnarray}

\begin{eqnarray}
F^{(1)}_{X}&=&1-\left(\frac{45}{4}+\frac{15}{4} \text{Cos}[4 \alpha ]-\frac{15}{2} \text{Cos}[2 \beta ] \text{Sin}[2 \alpha ]^2\right) p_x-\left(\frac{69}{4}-\frac{9}{4} \text{Cos}[4 \alpha ]+\frac{3}{2} \text{Cos}[2 \beta ] \text{Sin}[2 \alpha ]^2\right) p_y\nonumber\\
&-&\left(\frac{21}{2}-\frac{21}{2} \text{Cos}[4 \alpha ]\right) p_z+\left(\frac{315}{2}+\frac{105}{2} \text{Cos}[4 \alpha ]-105 \text{Cos}[2 \beta ] \text{Sin}[2 \alpha ]^2\right) p_x^2\nonumber\\
&+&\left(342+78 \text{Cos}[4 \alpha ]-264 \text{Cos}[2 \beta ] \text{Sin}[2 \alpha ]^2\right) p_x p_y+\left(\frac{573}{2}-\frac{153}{2} \text{Cos}[4 \alpha ]+57 \text{Cos}[2 \beta ] \text{Sin}[2 \alpha ]^2\right) p_y^2\nonumber\\
&+&\left(153-153 \text{Cos}[4 \alpha ]-306 \text{Cos}[2 \beta ] \text{Sin}[2 \alpha ]^2\right) p_x p_z+\left(393-393 \text{Cos}[4 \alpha ]+54 \text{Cos}[2 \beta ] \text{Sin}[2 \alpha ]^2\right) p_y p_z\nonumber\\
&+&(210-210 \text{Cos}[4 \alpha ]) p_z^2
\label{Not1qubitfidelity}
\end{eqnarray}

\begin{eqnarray}
F^{(7)}_{P}&=&1-51 p_x-57 p_y-(53-4 \text{Cos}[4 \alpha ]) p_z+\left(1403-6 \text{Cos}[4 \alpha ]+12 \text{Cos}[2 \beta ] \text{Sin}[2 \alpha ]^2\right) p_x^2\nonumber\\
&+&\left(2928+10 \text{Cos}[4 \alpha ]-4 \text{Cos}[2 \beta ] \text{Sin}[2 \alpha ]^2\right) p_x p_y+\left(\frac{3259}{2}+\frac{3}{2} \text{Cos}[4 \alpha ]+\text{Cos}[2 \beta ] \text{Sin}[2 \alpha ]^2\right) p_y^2\nonumber\\
&+&(2656-200 \text{Cos}[4 \alpha ]) p_x p_z+(3026-218 \text{Cos}[4 \alpha ]) p_y p_z+(1581-190 \text{Cos}[4 \alpha ]) p_z^2
\label{Phase7qubitfidelity}
\end{eqnarray}

\begin{eqnarray}
F^{(1)}_{P}&=&1-\left(\frac{81}{4}-\frac{21}{4} \text{Cos}[4 \alpha ]+\frac{3}{2} \text{Cos}[2 \beta ] \text{Sin}[2 \alpha ]^2\right) p_x+\left(-\frac{61}{4}+\frac{1}{4} \text{Cos}[4 \alpha ]-\frac{9}{2} \text{Cos}[2 \beta ] \text{Sin}[2 \alpha ]^2\right) p_y\nonumber\\
&-&\left(\frac{23}{2}-\frac{23}{2} \text{Cos}[4 \alpha ]\right) p_z+\left(\frac{837}{2}-\frac{417}{2} \text{Cos}[4 \alpha ]+75 \text{Cos}[2 \beta ] \text{Sin}[2 \alpha ]^2\right) p_x^2\nonumber\\
&+&\left(603-183 \text{Cos}[4 \alpha ]+270 \text{Cos}[2 \beta ] \text{Sin}[2 \alpha ]^2\right) p_x p_y+\left(\frac{455}{2}-\frac{35}{2} \text{Cos}[4 \alpha ]+135 \text{Cos}[2 \beta ] \text{Sin}[2 \alpha ]^2\right) p_y^2\nonumber\\
&+&\left(570-570 \text{Cos}[4 \alpha ]+72 \text{Cos}[2 \beta ] \text{Sin}[2 \alpha ]^2\right) p_x p_z+\left(350-350 \text{Cos}[4 \alpha ]+204 \text{Cos}[2 \beta ] \text{Sin}[2 \alpha ]^2\right) p_y p_z\nonumber\\
&+&(253-253 \text{Cos}[4 \alpha ]) p_z^2.
\label{Phase1qubitfidelity}
\end{eqnarray}

Eq.~\ref{PQEC} represents the fidelity of an arbitrary state that has undergone error-prone encoding, followed 
by a perfect (non error-prone) quantum error correction scheme. Note that the only first order error probability term 
is $p_{z}$ (and this term falls out for $\alpha = 0,\frac{\pi}{2}$):

\begin{eqnarray}
F_{P-QEC,enc}^{(7)}&=&1-(2-2 \text{Cos}[4 \alpha ]) p_z-\left(\frac{9}{2}+\frac{3}{2} \text{Cos}[4 \alpha ]-3 \text{Cos}[2 \beta ] \text{Sin}[2 \alpha ]^2\right) p_x^2\nonumber\\
&-&\left(6-2 \text{Cos}[4 \alpha ]+2 \text{Cos}[2 \beta ] \text{Sin}[2 \alpha ]^2\right) p_x p_y-\left(\frac{3}{4}+\frac{1}{4} \text{Cos}[4 \alpha ]-\frac{1}{2} \text{Cos}[2 \beta ] \text{Sin}[2 \alpha ]^2\right) p_y^2\nonumber\\
&-&(2-2 \text{Cos}[4 \alpha ]) p_x p_z-\left(\frac{7}{2}-\frac{7}{2} \text{Cos}[4 \alpha ]\right) p_y p_z-\left(\frac{3}{2}-\frac{3}{2} \text{Cos}[4 \alpha ]\right) p_z^2.
\label{PQEC}
\end{eqnarray}

Eqs.~\ref{PQECHad}, \ref{PQECNot}, and \ref{PQECPhase} represent the fidelities of an arbitrary state that has 
undergone error-prone encoding and error-prone application of the logical single qubit Clifford gates (Hadamard, 
\textsc{not}, and phase, respectively), followed by perfect quantum error correction. Note that the only first order 
term in all three equations is $p_{z}$ (and this term falls out for $\alpha = 0,\frac{\pi}{2}$):

\begin{eqnarray}
F_{P-QEC,H}^{(7)}&=&1-(2-2 \text{Cos}[4 \alpha ]) p_z-\left(\frac{9}{2}+\frac{3}{2} \text{Cos}[4 \alpha ]-3 \text{Cos}[2 \beta ] \text{Sin}[2 \alpha ]^2\right) p_x^2-\left(6-2 \text{Cos}[4 \alpha ]+2 \text{Cos}[2 \beta ] \text{Sin}[2 \alpha ]^2\right) p_x p_y\nonumber\\
&-&\left(\frac{3}{4}+\frac{1}{4} \text{Cos}[4 \alpha ]-\frac{1}{2} \text{Cos}[2 \beta ] \text{Sin}[2 \alpha ]^2\right) p_y^2-\left(7-3 \text{Cos}[4 \alpha ]-2 \text{Cos}[2 \beta ] \text{Sin}[2 \alpha ]^2\right) p_x p_z\nonumber\\
&-&\left(\frac{23}{4}-\frac{19}{4} \text{Cos}[4 \alpha ]+\frac{1}{2} \text{Cos}[2 \beta ] \text{Sin}[2 \alpha ]^2\right) p_y p_z-\left(\frac{3}{2}-\frac{3}{2} \text{Cos}[4 \alpha ]\right) p_z^2
\label{PQECHad}
\end{eqnarray}

\begin{eqnarray}
F_{P-QEC,X}^{(7)}&=&1-(2-2 \text{Cos}[4 \alpha ]) p_z-\left(\frac{27}{4}+\frac{9}{4} \text{Cos}[4 \alpha ]-\frac{9}{2} \text{Cos}[2 \beta ] \text{Sin}[2 \alpha ]^2\right) p_x^2\nonumber\\
&-&\left(\frac{13}{2}-\frac{5}{2} \text{Cos}[4 \alpha ]+2 \text{Cos}[2 \beta ] \text{Sin}[2 \alpha ]^2\right) p_x p_y-\left(\frac{3}{4}+\frac{1}{4} \text{Cos}[4 \alpha ]-\frac{1}{2} \text{Cos}[2 \beta ] \text{Sin}[2 \alpha ]^2\right) p_y^2\nonumber\\
&-&(2-2 \text{Cos}[4 \alpha ]) p_x p_z-\left(\frac{7}{2}-\frac{7}{2} \text{Cos}[4 \alpha ]\right) p_y p_z-\left(\frac{7}{2}-\frac{7}{2} \text{Cos}[4 \alpha ]\right) p_z^2
\label{PQECNot}
\end{eqnarray}

\begin{eqnarray}
F_{P-QEC,P}^{(7)}&=&1-(2-2 \text{Cos}[4 \alpha ]) p_z-\left(\frac{9}{2}+\frac{3}{2} \text{Cos}[4 \alpha ]-3 \text{Cos}[2 \beta ] \text{Sin}[2 \alpha ]^2\right) p_x^2-(9-\text{Cos}[4 \alpha ]) p_x p_y\nonumber\\
&-&(3-\text{Cos}[4 \alpha ]) p_y^2-(2-2 \text{Cos}[4 \alpha ]) p_x p_z-\left(\frac{7}{2}-\frac{7}{2} \text{Cos}[4 \alpha ]\right) p_y p_z-\left(\frac{7}{2}-\frac{7}{2} \text{Cos}[4 \alpha ]\right) p_z^2.
\label{PQECPhase}
\end{eqnarray}

Eq. \ref{QEC} represents the fidelity of an arbitrary state that has been encoded via the error-prone 
$[7,1,3]$ code and has then undergone error-prone quantum error correction. Eq. \ref{QEC1} represents the fidelity of the single logical qubit of this state:

\begin{eqnarray}
F_{QEC}^{(7)}&=&1-55 p_x-7 p_y-(9-2 \text{Cos}[4 \alpha ]) p_z+\left(\frac{3111}{2}-\frac{3}{2} \text{Cos}[4 \alpha ]+3 \text{Cos}[2 \beta ] \text{Sin}[2 \alpha ]^2\right) p_x^2\nonumber\\
&+&\left(226+2 \text{Cos}[4 \alpha ]-2 \text{Cos}[2 \beta ] \text{Sin}[2 \alpha ]^2\right) p_x p_y-\left(\frac{1079}{4}+\frac{1}{4} \text{Cos}[4 \alpha ]-\frac{1}{2} \text{Cos}[2 \beta ] \text{Sin}[2 \alpha ]^2\right) p_y^2\nonumber\\
&+&(330-103 \text{Cos}[4 \alpha ]) p_x p_z-\left(\frac{1293}{2}+\frac{21}{2} \text{Cos}[4 \alpha ]\right) p_y p_z-\left(\frac{297}{2}+\frac{5}{2} \text{Cos}[4 \alpha ]\right) p_z^2
\label{QEC}
\end{eqnarray}

\begin{eqnarray}
F_{QEC}^{(1)}&=&1-\left(\frac{57}{4}+\frac{19}{4} \text{Cos}[4 \alpha ]-\frac{19}{2}\text{Cos}[2\beta]\text{Sin}[2\alpha]^{2}\right) p_x-\left(\frac{13}{4}-\frac{1}{4} \text{Cos}[4 \alpha ]-\frac{1}{2}\text{Cos}[2\beta]\text{Sin}[2\alpha]^{2}\right) p_y \nonumber\\
&-&\left(\frac{7}{2}-\frac{7}{2} \text{Cos}[4 \alpha ]\right) p_z +(210+70 \text{Cos}[4 \alpha ]-140 \text{Cos}[2\beta]\text{Sin}[2\alpha]^{2}) p_x^2\nonumber\\
&+&\left(\frac{169}{4}+\frac{43}{4} \text{Cos}[4 \alpha ]-\frac{109}{2}\text{Cos}[2\beta]\text{Sin}[2\alpha]^{2}\right) p_x p_y-\left(\frac{175}{2}+\frac{37}{2} \text{Cos}[4 \alpha ]-34\text{Cos}[2\beta]\text{Sin}[2\alpha]^{2}\right) p_y^2\nonumber\\
&+&\left(\frac{51}{4}-\frac{271}{4} \text{Cos}[4 \alpha ]-\frac{209}{2}\text{Cos}[2\beta]\text{Sin}[2\alpha]^{2}\right) p_x p_z-\left(\frac{829}{4}+\frac{127}{4} \text{Cos}[4 \alpha ]-\frac{137}{2}\text{Cos}[2\beta]\text{Sin}[2\alpha]^{2}\right) p_y p_z\nonumber\\
&-&(42-42 \text{Cos}[4 \alpha ]) p_z^2.
\label{QEC1}
\end{eqnarray}

Eq. \ref{QEC2X} represents the fidelity of an arbitrary state that has been encoded via the error-prone 
$[7,1,3]$ code and has undergone error-prone quantum error correction two times successively. Eq. \ref{QEC2X1} represents the fidelity of the single logical qubit of this procedure:

\begin{eqnarray}
F_{2xQEC}^{(7)}&=&1-55 p_x-7 p_y-(9-2 \text{Cos}[4 \alpha ]) p_z+\left(\frac{3081}{2}-\frac{3}{2} \text{Cos}[4 \alpha ]+3 \text{Cos}[2 \beta ] \text{Sin}[2 \alpha ]^2\right) p_x^2\nonumber\\
&+&\left(229+2 \text{Cos}[4 \alpha ]-2 \text{Cos}[2 \beta ] \text{Sin}[2 \alpha ]^2\right) p_x p_y-\left(\frac{1059}{4}+\frac{1}{4} \text{Cos}[4 \alpha ]-\frac{1}{2} \text{Cos}[2 \beta ] \text{Sin}[2 \alpha ]^2\right) p_y^2\nonumber\\
&+&(330-103 \text{Cos}[4 \alpha ]) p_x p_z-\left(\frac{1235}{2}+\frac{21}{2} \text{Cos}[4 \alpha ]\right) p_y p_z-\left(\frac{271}{2}+\frac{5}{2} \text{Cos}[4 \alpha ]\right) p_z^2
\label{QEC2X}
\end{eqnarray}

\begin{eqnarray}
F_{2xQEC}^{(1)}&=&1-\left(\frac{57}{4}+\frac{19}{4} \text{Cos}[4 \alpha ]-\frac{19}{2}\text{Cos}[2 \beta ] \text{Sin}[2 \alpha ]^2\right) p_x-\left(\frac{13}{4}-\frac{1}{4} \text{Cos}[4 \alpha ]-\frac{1}{2}\text{Cos}[2 \beta ] \text{Sin}[2 \alpha ]^2\right) p_y\nonumber\\
&-&\left(\frac{7}{2}-\frac{7}{2} \text{Cos}[4 \alpha ]\right) p_z+(198+66 \text{Cos}[4 \alpha ]-132\text{Cos}[2 \beta ] \text{Sin}[2 \alpha ]^2) p_x^2\nonumber\\
&+&\left(\frac{177}{4}+\frac{43}{4} \text{Cos}[4 \alpha ]-\frac{107}{2}\text{Cos}[2 \beta ] \text{Sin}[2 \alpha ]^2\right) p_x p_y-\left(85+19 \text{Cos}[4 \alpha ]-33\text{Cos}[2 \beta ] \text{Sin}[2 \alpha ]^2\right) p_y^2\nonumber\\
&+&\left(\frac{51}{4}-\frac{271}{4} \text{Cos}[4 \alpha ]-\frac{209}{2}\text{Cos}[2 \beta ] \text{Sin}[2 \alpha ]^2\right) p_x p_z-\left(\frac{773}{4}+\frac{135}{4} \text{Cos}[4 \alpha ]-\frac{133}{2}\text{Cos}[2 \beta ] \text{Sin}[2 \alpha ]^2\right) p_y p_z\nonumber\\
&-&(38-38 \text{Cos}[4 \alpha ]) p_z^2
\label{QEC2X1}
\end{eqnarray}

Eq. \ref{QECNot} represents the fidelity of an arbitrary state that has been encoded via the error-prone 
$[7,1,3]$ code, and has then undergone noisy application of the logical single qubit NOT gate, followed 
by error-prone quantum error correction. Eq. \ref{QECNot1} represents the fidelity of the single logical qubit of this procedure:

\begin{eqnarray}
F_{QEC,X}^{(7)}&=&1-55 p_x-7 p_y-(9-2 \text{Cos}[4 \alpha ]) p_z+\left(\frac{6193}{4}-\frac{9}{4} \text{Cos}[4 \alpha ]+\frac{9}{2} \text{Cos}[2 \beta ] \text{Sin}[2 \alpha ]^2\right) p_x^2\nonumber\\
&+&\left(\frac{451}{2}+\frac{5}{2} \text{Cos}[4 \alpha ]-2 \text{Cos}[2 \beta ] \text{Sin}[2 \alpha ]^2\right) p_x p_y-\left(\frac{1079}{4}+\frac{1}{4} \text{Cos}[4 \alpha ]-\frac{1}{2} \text{Cos}[2 \beta ] \text{Sin}[2 \alpha ]^2\right) p_y^2\nonumber\\
&+&(330-103 \text{Cos}[4 \alpha ]) p_x p_z-\left(\frac{1309}{2}+\frac{21}{2} \text{Cos}[4 \alpha ]\right) p_y p_z-\left(\frac{311}{2}+\frac{1}{2} \text{Cos}[4 \alpha ]\right) p_z^2
\label{QECNot}
\end{eqnarray}

\begin{eqnarray}
F_{QEC,X}^{(1)}&=&1-\left(\frac{57}{4}+\frac{19}{4} \text{Cos}[4 \alpha ]-\frac{19}{2}\text{Cos}[2 \beta ] \text{Sin}[2 \alpha ]^2\right) p_x-\left(\frac{13}{4}-\frac{1}{4} \text{Cos}[4 \alpha ]-\frac{1}{2}\text{Cos}[2 \beta ] \text{Sin}[2 \alpha ]^2\right) p_y\nonumber\\
&-&\left(\frac{7}{2}-\frac{7}{2} \text{Cos}[4 \alpha ]\right) p_z+\left(\frac{819}{4}+\frac{273}{4} \text{Cos}[4 \alpha ]-\frac{273}{2}\text{Cos}[2 \beta ] \text{Sin}[2 \alpha ]^2\right) p_x^2\nonumber\\
&+&\left(\frac{167}{4}+\frac{45}{4} \text{Cos}[4 \alpha ]-\frac{109}{2}\text{Cos}[2 \beta ] \text{Sin}[2 \alpha ]^2\right) p_x p_y-\left(\frac{175}{2}+\frac{37}{2} \text{Cos}[4 \alpha ]-34\text{Cos}[2 \beta ] \text{Sin}[2 \alpha ]^2\right) p_y^2\nonumber\\
&+&\left(\frac{51}{4}-\frac{271}{4} \text{Cos}[4 \alpha ]-\frac{209}{2}\text{Cos}[2 \beta ] \text{Sin}[2 \alpha ]^2\right) p_x p_z-\left(\frac{847}{4}+\frac{125}{4} \text{Cos}[4 \alpha ]-\frac{139}{2}\text{Cos}[2 \beta ] \text{Sin}[2 \alpha ]^2\right) p_y p_z\nonumber\\
&-&(46-46 \text{Cos}[4 \alpha ]) p_z^2.
\label{QECNot1}
\end{eqnarray}

\end{widetext}

\end{document}